\documentclass[12pt,preprint]{aastex}

\slugcomment{Not to appear in Nonlearned J., 45.}

\shorttitle{TeV neutrinos from microquasars}
\shortauthors{W. Bednarek}

\begin{document}


\title{TeV neutrinos from microquasars in compact massive binaries}


\author{W\l odek Bednarek\altaffilmark{1}}
%\affil{Department of Experimental Physics, University of \L \'od\'z,
%ul. Pomorska 149/153, 90-236 \L \'od\'z, Poland}

\altaffiltext{1}{Department of Experimental Physics, University of \L \'od\'z,
ul. Pomorska 149/153, 90-236 \L \'od\'z, Poland; bednar@fizwe4.fic.uni.lodz.pl}

\begin{abstract}
We consider a compact binary system in which a Wolf-Rayet star supplies 
matter onto a stellar mass black hole or a neutron star. This matter forms an
accretion disk which ejects a jet as observed in Galactic microquasars. 
A part of the jet kinetic energy, typically $\sim 10\%$, 
can be transfered to relativistic nuclei. These 
nuclei lose nucleons as a result of photo-disintegration process in collisions with 
thermal photons from the accretion disk and the massive
star. Due to the head on photon-nucleus collisions most of neutrons released from 
nuclei move towards the surface of the accretion disk and/or the massive star 
producing neutrinos in collisions with the matter. We calculate the spectra of muon
neutrinos and expected neutrino event rates in a 1 km$^3$ neutrino detector of the 
IceCube type from a microquasar inside our Galaxy applying, as an example, the 
parameters of the Cyg X-3 binary system. Such model predicts up to several neutrino 
events per km$^3$ per yr from Cyg X-3, provided that nuclei are accelerated 
to the Lorentz factors above $10^6$ with a power law spectrum with an index 
close to 2.
\end{abstract}


\section{Introduction}

A part of X-ray binary systems, called microquasars, produce collimated outflows 
observed as 
nonthermal radio structures (e.g. Mirabel \& Rodriquez~1999, Fender~2003).
They are probably produced by relativistic jets launched from the inner parts of the 
accretion disks around stellar mass black holes or neutron 
stars. High energy processes play an important
role in these sources  since some of them emit high energy X-rays and $\gamma$-rays 
(e.g. Paredes~2005).
However, the content of these jets, hadronic, leptonic or mixed, remains a puzzle.
A few models have been discussed in which possible importance of hadronic processes
in the microquasar jets can be probed by the neutrino observations.
For example, Levinson \& Waxman~(2001) and Distefano et al.~(2002) consider 
production of neutrinos in the interactions of protons accelerated in the jet with 
the radiation field of the accretion disk.
Romero et al.~(2003) discuss the case of a misaligned jet, which 
collide with dense wind of a massive star or its surface. Aharonian \& Atoyan~(1991)
propose that relativistic hadronic beam can interact with blobs of matter 
injected by the stellar companion.
Other scenarios for neutrino production in X-ray binaries, which do not postulate
very collimated injection of relativistic hadrons,  
have been also discussed, e.g. collisions of hadrons from the pulsar with matter 
of the accretion disk (Cheng \& Ruderman 1989, Anchordoqui et al. 2003) or the matter 
of the massive star (Gaisser \& Stanev~1985, Kolb et al.~1985, Berezinsky et al.~1985,
Harding \& Gaisser~1990).

In this paper we consider a different model for the neutrino production in 
microquasars in which the jet does not need to collide directly with the dense matter 
from the stellar companion (a rather rare phenomenon due to the geometrical
requirements). We show that in very compact massive binaries containing the massive
star and a compact object, e.g. the Cyg X-3, nuclei 
accelerated in the jet can lose efficiently neutrons in collisions with the stellar 
and accretion disk radiation. Most of these neutrons propagate towards the disk and 
the massive star producing neutrinos in collisions with the matter. 
Note that about half of the observed Wolf-Rayet stars, between 158 objects
(van der Hucht~2001), form massive binary stars. Such systems 
should soon evolve into the binary system containing a compact object, i.e. 
similar to the Cyg X-3. In fact, a few high mass compact X-ray binaries with
the solar mass black hole are observed at present, e.g. Cyg X-1, SS 433, V4641 Sgr 
(Paredes 2005).

\section{A compact object close to a massive star}

Let's consider a solar mass black hole or a slowly rotating neutron star on a close
orbit around a Wolf-Rayet (WR) star. 
The matter accreting from the massive star forms an accretion 
disk which launches a jet from its inner part as observed in microquasars.
As an example, we consider a specific case of the Cyg X-3 binary system 
which basic parameters are the following: the Wolf-Rayet star 
with the surface temperature $T_{\rm WR} = 1.36\times 10^5$ K and radius 
$R_{\rm WR} = 1.6 R_\odot$, the separation of the components $D = (3.6\pm 1.2)R_\odot
= 2.25 R_{\rm WR}$. Persistent 
X-ray emission of the accretion disk in Cyg X-3 is 
$L_{\rm d}\approx 10^{38}$ erg s$^{-1}$ (Bonnet-Bidaud \& Chardin~1988).
For this disk luminosity, and an assumed inner radius of the disk $R_{\rm d} = 10^7$ cm, 
the inner disk temperature is $T_{\rm d}\sim 10^7$ K. According to the the disk-jet 
symbiosis model (Falcke, Malkan \& Biermann~1995), the jet power, 
$L_{\rm j}$, is comparable to the observed disk luminosity, $L_{\rm d}$
(this model has been recently successfully applied for different
sources, see e.g. Yuan et al. 2002, Yuan, Markoff \& Falcke 2002, Falcke et al. 2004). 
Therefore, in farther considerations we assume 
$L_{\rm j} = 0.5L_{\rm d} = 5\times 10^{37}$ erg s$^{-1}$.
Note however, that in the outburst stages, which are relatively frequent in Cyg X-3, 
the X-ray luminosity can be several times larger. The jet in Cyg X-3 propagates with 
the velocity in the range (0.3-0.8)c ($c$ is the velocity of light), estimated 
from the expension rate of specific components (e.g. Spencer et al.~1986) and the 
ratio of the jet to counter-jet flux (Mioduszewski et al.~2001).  
We apply the average value, $v_{\rm j} = 0.5c$, reported by Marti et al.~(2001).

The WR stars are at the late stage of evolution of the supermassive stars which have 
already lost their hydrogen envelopes. Their winds are mainly composed
from the CNO group of nuclei with the substantial abundance of He nuclei
(see e.g. Heger \& Langer~2000). We assume that these nuclei can be accelerated in 
some regions of the jet to relativistic energies. They create a relatively weakly
relativistic blob in which nuclei with very large Lorentz factors are distributed
isotropically in the blob reference frame. Those among these nuclei which happen to 
move in the direction of the source of soft photons, i.e. the inner part of the 
accretion disk or the disk of the massive star, suffer the most frequent 
disintegrations. As a result, most of neutrons extracted from these nuclei move along
the straight lines towards the source of the radiation field. 
The number of neutrons which move towards the disk and the star is determined by 
the probability of neutron extraction from nuclei in such strongly anisotropic 
radiation defined by $\int A(\theta) f d(\cos\theta)$, 
where $A(\theta)$ is the product of the cross section and the photon spectrum,
$f = 1 + \cos\theta$ and $\theta$ is the angle between the photon and nuclei.
Let us assume that $A(\theta)$ weakly depends on $\theta$, i.e. assuming that
the interaction regime is far away from the threshold for extraction of 
a nucleon from nuclei. Then we estimate that 
the probability of interaction of a nucleus which move inside the half hemisphere 
around the source of soft radiation is at least a factor 3 larger than for nucleus
moving in the opposite half of the hemisphere. Therefore, at least
3 times more neutrons, extracted from nuclei, move towards the accretion disk and/or 
the massive star than at other directions. Note that at low energies the 
photo-disintegration cross section drops and also depends on $f$. Then
significantly larger fraction of 
neutrons extracted from nuclei should move towards the source of the soft photons.

The location 
of the acceleration place in the jet, the dissipation region, is unknown and depends
on the details of the acceleration model. Therefore, we consider three different
possible regions inside the jet which location is determined by the strength of 
the radiation field from the disk and the massive star:

(I) Nuclei accelerated to sufficient energies very close to the base of the jet are 
immediately disintegrated in collisions with X-ray photons from the inner 
accretion disk. Nucleons from their disintegration may interact with the  
X-ray photons producing neutrinos through decay of pions, provided that they have
Lorentz factors above the threshold for pion production, $\gamma^{\rm min}_\pi\approx
m_\pi/6k_{\rm B}T_{\rm d}\approx  10^5$, where $m_\pi$ is the pion mass and 
$k_{\rm B}$ is the Boltzmann constant.
Such scenario is realized if the optical depth for pion production by nucleons in  
collisions with the disk photons is above unity. 
The extent of the region (I) can be estimated 
by comparing the characteristic escape time 
of nucleons with the jet plasma, $\tau_{\rm esc} = xR_{\rm d}/v_{\rm j}$,
where $x$ is the distance along the jet in units of the inner disk radius 
$R_{\rm d}$, with the interaction 
time of nucleons $\tau_{N\varepsilon\rightarrow\pi} =  
(n_{\rm d}\sigma_{N\varepsilon}c)^{-1}$,
where $\sigma_{N\varepsilon} = 
(1-4)\times 10^{-28}$ cm$^2$ is the cross section for proton-photon pion production 
in the region of the plateau and the peak, respectively. 
The photon density from the inner disk is approximated by  
\begin{eqnarray}
n_{\rm d}\approx {{4\sigma_{\rm SB}T_{\rm d}^4}\over{3ck_{\rm B}T_{\rm d}x^2}}\approx
{{2\times 10^{16}T_5^3}\over{x^2}}~~~{\rm ph.~cm^{-3}}
\label{eq1}
\end{eqnarray}
\noindent
where $T_{\rm d} = 10^5 T_5$ is the disk temperature at its inner radius,
$\sigma_{\rm SB}$ is the Stefan-Boltzmann constant. Region (I) extends up to
$x_{\rm N\varepsilon\rightarrow\pi}\approx (40-170)$, i.e. at 
$\sim (4-17)\times 10^8$
cm  from the base of the jet (see Fig.~1). We do not consider here the region (I) 
since it 
has been discussed recently by Levinson \& Waxman~(2001) and Distefano et al.~(2002).

(II) Nucleons from disintegration of nuclei are not able to produce efficiently 
pions in collisions with the disk radiation but they are able to fragment 
significantly on separate nucleons. Due to the head on collisions of the nuclei 
with photons, neutrons move towards the inner disk producing 
neutrinos in collisions with the matter. Such general picture 
(with the difference of injection of neutrons by protons) has been proposed for the 
accretion disks in active galactic nuclei (Nellen et al~1993). 
The region (II) extends along the jet to the distance, $x_{\rm d/s}$, at which 
density of photons from the disk becomes comparable to that 
from the WR star (see Fig.~1). By comparing these photon densities, we estimate 
$x_{\rm d/s}\approx 10^3$. The photon density from the 
massive star at the distance equal to the separation of the components is 
$n_{\rm WR}\approx 
4\sigma_{\rm SB}T_{\rm WR}^4/(3ck_{\rm B}T_{\rm WR})(R_{\rm WR}/D)^2$.
For density of photons from the disk see Eq.~\ref{eq1}.

(III) In this region radiation from the WR star dominates. As we show later
this region is at least an order of magnitude larger
than the inner regions dominated by the radiation of the accretion disk. 
Neutrons extracted from nuclei in the region (III) move mainly towards the surface 
of the WR star due to the head on collisions of nuclei with the stellar photons.

In conclusion, depending on the region in the jet where nuclei are accelerated
above the threshold for their efficient photo-disintegrations (I, II, or III), we 
expect production of neutrinos either in the direction along the jet (region I), 
or in the direction opposite to the jet propagation, i.e. through the accretion disk 
(region II), or at the wide solid angle centered on the plane of the binary system 
(region III).  
Note, that in the case of significant Doppler boosting the borders between specific 
regions move down, since the disk radiation plays minor role due to the collimation 
of the relativistic nuclei in the jet. This effect is not considered here in detail
since the Doppler factor of the jet in Cyg X-3 is not far from unity. 

\section{Relativistic nuclei in the jet}

In order to estimate the maximum energies of nuclei accelerated in the jet, let us 
assume that the magnetic field energy density in the inner part of the accretion disk 
(at the base of the jet) is in the equipartition with the energy density of radiation,
i.e. $U_{\rm B}\approx U_{\rm d}$. For the disk temperature at the inner radius,
$T_{\rm d} = 10^7$ K, the magnetic field can be as large as $B_{\rm d}\approx 3\times 10^7$ G. 
The magnetic field drops along the jet according to $B(x) =
B_{\rm d}/(1 +\alpha x)\approx B_{\rm d}/x$ (for $x\gg 1/\alpha$),
where $\alpha = 0.1$ rad is the jet opening angle.
The acceleration rate of nuclei can be expressed by 
\begin{eqnarray}
d\gamma_{\rm A}/dt = \xi ZeBc/Am_{\rm p}, 
\label{eq2}
\end{eqnarray}
\noindent
where $\xi$ is the efficiency of acceleration mechanism, $Am_p$, and $Ze$
are the mass and the charge of the nuclei, and $Z/A = 0.5$. Nuclei can in principle
reach the maximum Lorentz factors allowed by the spatial extend of 
the acceleration region, $L_{\rm acc}\approx \alpha x R_{\rm d}$, as large as
\begin{eqnarray}
\gamma_{\rm A}^{\rm max}\approx \alpha x R_{\rm d}(d\gamma_{\rm A}/dt)/c = 
\xi e B_{\rm d}R_{\rm d}\approx 10^8\xi. 
\label{eq3}
\end{eqnarray}
\noindent
The maximum Lorentz factors of nuclei can be limited by their escape time along 
the jet, $\tau_{\rm esc}$ (estimated above), or by their energy 
losses in collisions with thermal photons on the pion and the disintegration 
processes (the energy losses on $e^\pm$ production can be neglected in respect to
these processes). By comparing the acceleration time scale of nuclei in the jet,
$\tau_{\rm acc} = \gamma_{\rm A}/(d\gamma_{\rm A}/dt)$ with their escape time scale,
$\tau_{\rm esc}$,
we conclude that nuclei can be potentially accelerated in the jet up to 
$\gamma_{\rm A}^{\rm max}$ (given by Eq.~\ref{eq3}). It has been shown above that
in the regions II and III of the jet, the pion interaction time scale is longer than 
the escape time scale and so the acceleration time scale. Therefore, the acceleration 
of nuclei in regions II and III of the jet is not limited by their energy losses 
on pion production. Below we discuss in detail the role of disintegration 
process in our model.

Heavy relativistic nuclei can lose efficiently nucleons due to the 
photo-disintegration process in collisions with thermal photons from the 
accretion disk or the massive star if  
photon energies in the reference frame of the nuclei are above $\sim 10$ MeV, i.e. 
$E_\gamma = 6k_BT\gamma_A = 5\times 10^{-5}T_5\gamma_{\rm A}~~{\rm MeV} > 
10~{\rm MeV}$. 
This condition gives the lower limit on the Lorentz factor of accelerated 
nuclei above which their photo-disintegration becomes important, 
\begin{eqnarray}
\gamma_{A}^{min}\approx 2\times 10^5/T_5. 
\label{eq5}
\end{eqnarray}
The most efficient disintegrations occur when the photon energies correspond to the 
maximum in the photo-disintegration cross-section of
nuclei, which is  at $\sim 20$ MeV (with the half width of $\sim 10$ MeV) 
\cite{bi57}, for nuclei with the mass numbers between helium and oxygen. 

Let's estimate the efficiency of the photo-disintegration process of nuclei 
in the region II and III.
The mean free path for dissociation of a single nucleon from the nuclei is
$\lambda_{A\gamma} = (n\sigma_{A\gamma})^{-1}$, 
where $n$ is the density of thermal photons coming from the accretion disk
(see Eq.~\ref{eq1}), or the WR star, 
$n_{\rm WR}(r)\approx 4\sigma_{\rm SB}T_{\rm WR}^3/3ck_{\rm B}r^2$, where 
$r = R/R_{\rm WR}$, and $R$ is the distance from the WR star.
The photo-disintegration cross section at the peak of the giant resonance
and the plateau region after the resonance are:  
$\sigma_{\rm A\gamma}^{\rm r}\approx 1.45\times 10^{-27}A$ cm$^2$
(valid for $\gamma_{\rm A}\le 3\times 10^6$), and $\sigma_{\rm A \gamma}^{\rm p}\approx 
1.25\times 10^{-28}A$ cm$^2$ ($\gamma_{\rm A}>3\times 10^6$) \cite{kt93}.
The characteristic photo-disintegration time scales of nuclei with the mass number $A$
in the radiation field of the accretion disk ($T_{\rm d} = 10^7$ K) and the WR star
($T_{\rm WR}=1.36\times 10^5$ K), $\tau_{\rm A\gamma} = \lambda_{\rm A\gamma}/c$, are 
$\tau_{\rm A\gamma}^{\rm d}\approx (2.5-20)\times 10^{-6}x^2/A$ s 
and $\tau_{\rm A\gamma}^{\rm WR}\sim (0.5-4)r^2/A$ s, respectively. On the other site,
the convection escape time of nuclei along the jet is of the order of 
$\tau_{c}\approx \sqrt{R^2-D^2}/v_{\rm j}$, equal to $\sim 8 r$ s (if $R\gg D$).
By comparing $\tau_{\gamma A}^{\rm WR}$ and $\tau_{c}$, we get the upper limit for
the 
distance from the Wolf-Rayet star, $r_{\rm max}\approx (2-15)A$, below which the 
photo-disintegration process of nuclei is important. This is the upper bound on the 
region (III). Nuclei accelerated in the region (III) lose neutrons which fall onto 
the Wolf-Rayet star up at angles $\beta$ (measured from the plane of the binary 
system) which fulfill the condition: $\cos\beta = D/(r_{\rm max}R_{\rm WR})\approx 
(0.15-1)/A$. 

In order to determine the conditions for acceleration of nuclei let us compare
the acceleration time scale, $\tau_{\rm acc} = \gamma_{\rm A}/(d\gamma_{\rm A}/dt)
\approx x \gamma_{\rm A}/(3\times 10^{12}\xi)$ s, with the 
photo-disintegration time scale, $\tau_{\rm A\gamma}$. The maximum Lorentz factor to 
which nuclei can be accelerated before significant disintegration are then:
$\gamma_{dis}^{\rm II}\approx 7.5\times 10^6\xi x/A$ and 
$\gamma_{dis}^{\rm III} = 7.5\times 10^{12}\xi/(A x)$, 
for the region (II) and (III) respectively
(applying $r = 2.25$, $\alpha = 0.1$, and $A = 14$). The comparison of these Lorentz 
factors with the minimum Lorentz factor of nuclei for the
efficient photo-disintegration process (see Eq.~4) allows us to estimate the 
required value of the acceleration efficiency in the regions II and III, 
$\xi_{\rm th}^{\rm II}\approx 10^{-5}$ (for $x = 300$) and 
$\xi_{\rm th}^{\rm III}\approx 3\times 10^{-3}$ (for $x = 10^4$), above which nuclei 
at first obtain the power law spectrum and then lose separate neutrons. 
Neutrons separated from nuclei also have a power law spectrum with the spectral 
index of the parent nuclei. 
If $\xi < \xi_{\rm th}$, neutrons are injected by nuclei with the Lorentz factors 
close to $\gamma_{\rm A}^{\rm min}$, i.e. they are almost mono-energetic.
Therefore, we consider the injection of neutrons by nuclei with the mono-energetic and 
the power law spectra. 

\section{Production of neutrinos}

Most of neutrons extracted from nuclei in the region (II) 
move towards the inner part of the accretion disk and from the region (III)
towards the massive star producing 
neutrinos and $\gamma$-rays in hadronic interactions with the matter.
$\gamma$-rays, from the decay of neutral pions, are absorbed by very high column 
density of matter
(accept a narrow region around the stellar limb) but neutrinos can 
pass through the main part of the interior of the star without absorption. For the 
parameters of the Wolf-Rayet stars the optical depth is larger
than unity only for neutrinos moving close to the center of the star, i.e. with the
impact parameter less than $\sim 0.2 R_{\rm WR}$ which corresponds only to a few 
percent of the full stellar disk (see e.g. Bartosik et al.~2005).
Therefore, the fraction of neutrinos absorbed inside the star is relatively low
and can be neglected.
We calculate the muon neutrino (and anti-neutrino) spectra produced by neutrons
with the mono-energetic and the power law spectra. 
The multiple interactions of neutrons with the matter
have been taken into account up to their cooling to 100 GeV. 
We apply the scale breaking model for
hadronic interactions developed by Wdowczyk \& Wolfendale~(1987).
The neutrino spectra are calculated for the 
mono-energetic nuclei accelerated in the region (II) to the Lorentz factors 
$2\times 10^3$ and for the power law spectrum with the 
spectral indexes equal to 2 and 2.5 and the cut-offs at Lorentz factors $10^7$ and
$10^6$ (see Fig.~2a). We apply the distance to the Cyg X-3 binary system 
equal to 10 kpc and assume that $\chi = 10\%$ of the jet energy is transported to 
relativistic nuclei.
The spectra are clearly above the atmospheric neutrino background (ANB) provided 
that the spectrum of accelerated nuclei is flatter than 2.5 and extends  
above $\gamma_{\rm A} = 10^6$. The spectra produced by mono-energetic nuclei are 
significantly below the ANB. Therefore, acceleration of nuclei in the region II
with the efficiency $\xi < 10^{-5}$ can not be tested by neutrino telescopes.
The neutrino spectra produced by neutrons separated from nuclei accelerated in the 
region (III) in collisions with the matter of the massive star are shown in Fig.~2b. 
Neutrinos from regions II and III have similar fluxes.
They should be easily observed by the 1 km$^3$ neutrino detector of 
the IceCube type (Hill 2001). The smaller scale detectors can already
put interesting constraints on the efficiency of high energy processes in microquasars. 
For example, the case of mono-energetic injection of nuclei into the region (III) 
with the energy conversion efficiency $\chi = 10\%$ is already above the
upper limit on the neutrino flux from the Cyg X-3 derived on the base of the 
3 yr of data accumulated by the AMANDA II detector (Ackermann et al. 2004).
This limit is also within the factor of two to the predicted neutrino spectrum for 
the case of injection of nuclei with the power law spectrum and spectral index 2 
(see Figs.~2a and 2b). 

\section{Conclusion}

Let's estimate the neutrino event rate in a 1 km$^3$ neutrino detector 
from the Cyg X-3 type binary system. For that we apply the likelihood of neutrino 
detection by a 1 km$^3$ detector obtained 
by Gaisser \& Grillo~(1987) and the absorption coefficients of
neutrinos in the Earth obtained by Gandhi~(2000). We consider two limiting cases, 
i.e. the source close to the horizon (neutrinos not absorbed by the Earth), and the 
source at the nadir (neutrinos partially absorbed). The number of events
per year per km$^3$ expected for the mono-energetic injection of nuclei in the region 
III is $\sim 24$ events, and for nuclei with the power law spectrum 
in the region II (and the region III): 16.5 (14) events for the spectral index 
$\alpha = 2$ and the cut-off at the Lorentz factor $\gamma_{\rm A}^{max}=10^7$, and 
11 (8.3) events for the cut-off at $10^6$, and $\sim 2$ ($\sim 1$) events 
for $\alpha = 2.5$ and 
$\gamma_{\rm A}^{max}=10^7$. Since the neutrino spectra do not extend much
above a few $10^5$ GeV, the effects of absorption of neutrinos in the Earth are 
only of the order of $\sim 10\%$. 
Such event rates should be detected above the ANB
by the IceCube detector if the spectrum of accelerated nuclei is flatter than 2.5.
Note that predicted neutrino signals from the regions (II) and (III) should be emitted
in completely different directions. Neutrinos from the accretion disk are collimated
along the axis of the jet. In contrast, neutrinos from the WR star are emitted inside 
the broad solid angle around the plane of the binary system and should be
modulated with the orbital period of 
the binary system reaching the maximum when the compact object is behind the 
Wolf-Rayet star. This should help to separate the neutrino signal from the 
ANB. Therefore, even smaller scale neutrino detectors, e.g. of the AMANDA II and 
Antares type, might be also able to detect such microquasars in compact 
massive binaries within our Galaxy or at least put strong constraints on the efficiency 
of hadron acceleration in their jets.
The angular features of the neutrino emission, coupled with the information on the 
geometry of the binary system, should allow to distinguish between the two 
discussed regions for hadron acceleration giving information on the acceleration
process and formation of jets. 

The hadronic processes considered in this paper also turn to the production
of neutral pions which decay onto high energy gamma-rays.
Gamma-rays produced in the interaction of neutrons with the matter of 
the accretion disk initiate partial cascades which electromagnetic products 
might emerge from some parts of the accretion disk
(see e.g. Cheng \& Ruderman~1991). However, it has been shown
by e.g. Carraminana (1992) and Bednarek (1993) that gamma-rays with energies 
$\sim 1 - 10^3$ GeV should be strongly absorbed in the radiation field
produced by the accretion disk. Gamma-rays produced in 
the interaction of neutrons with the matter of the massive star are
absorbed inside the star except these ones which move
inside the narrow cone around the stellar
limb where the density of matter is low enough for the escape of
gamma-rays without significant absorption in the matter.
However these TeV gamma-rays are significantly absorbed in the thermal 
radiation of the massive star (Bednarek~2000, or Sierpowska \& Bednarek 2005)
since they are produced very close to the stellar surface.
We conclude that the hadronic model considered here does not predict 
gamma-ray fluxes which can be observed by the Cherenkov telescopes at energies 
above a few tens GeV. Only gamma-rays below these energies can escape from the 
dense soft radiation of the massive star and the accretion disk being potentially 
detected by the satellite experiments such as AGILE and GLAST.
However if significant amount of the jet energy is transfered to relativistic
leptons at relatively large distances from the base of the jet, then microquasars 
can also
produce observable fluxes of TeV gamma-rays in processes similar to that observed
in BL Lacs.

I would like to thank the anonymous referee for useful comments.
This work is supported by the KBN grants PBZ KBN054/P03/2001 and 1P03D01028.

\clearpage

\begin{figure}
\vskip 6.5truecm
\includegraphics{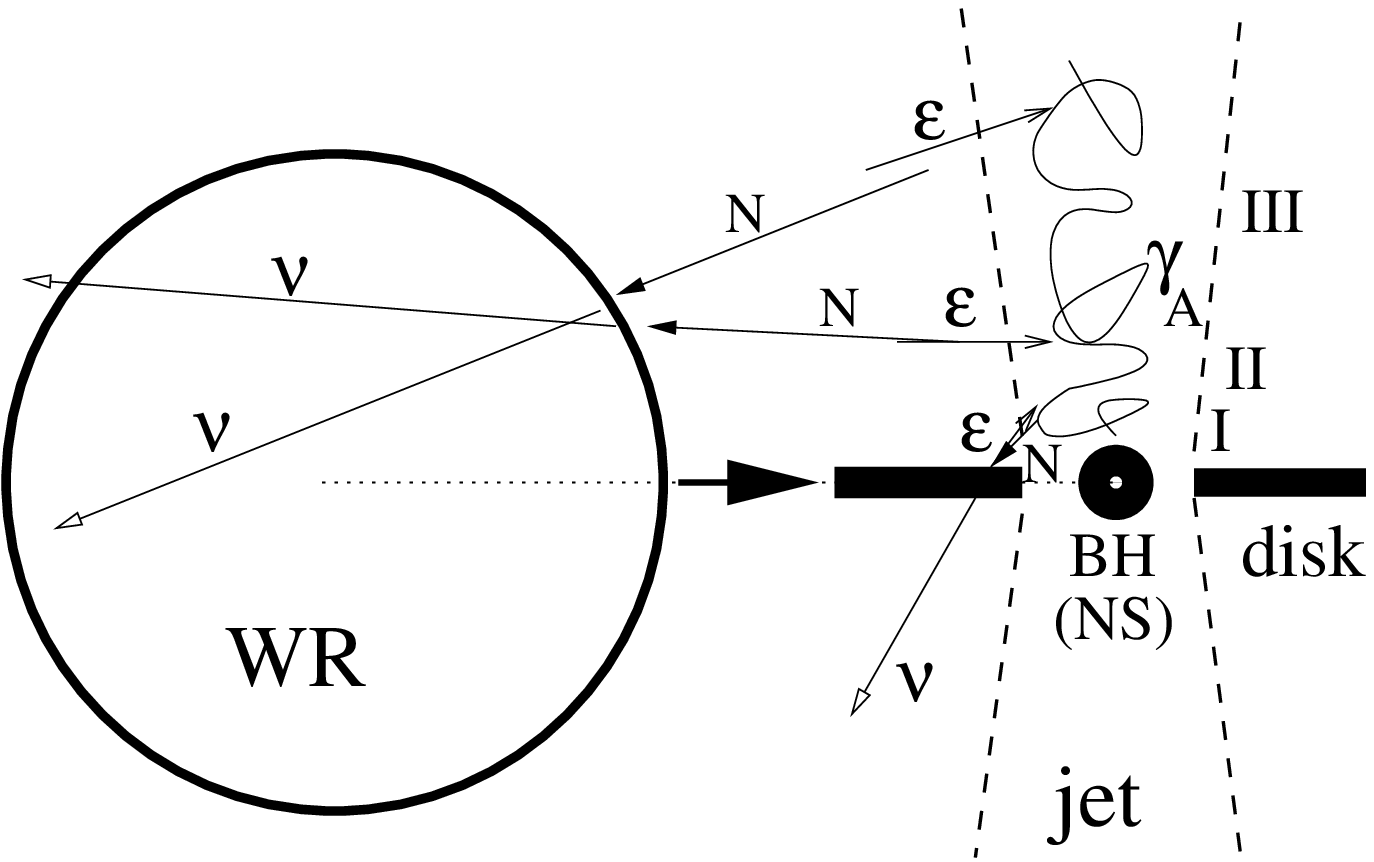}
\caption{\label{fig1}Schematic picture of a binary system containing 
a massive Wolf-Rayet star and a black hole or a neutron star (not to scale).
The matter from the massive star (mainly He and CNO nuclei) 
forms an accretion disk which launches a jet. Nuclei are accelerated at the jet up to 
the Lorentz factor $\gamma_{\rm A}$. If their Lorentz factors are  
$\gamma_{\rm A}>\gamma_{\rm A}^{min}$, they photo-disintegrate in collisions with 
thermal photons ($\varepsilon$) coming from the massive star and/or the inner part of 
the accretion disk losing separate nucleons.
Most of neutrons (N), separated from nuclei, move towards the massive star 
(due to the head on collisions) and/or to the inner part  of the accretion disk 
and interact with the matter producing neutrinos ($\nu$).}
\end{figure}
\begin{figure}
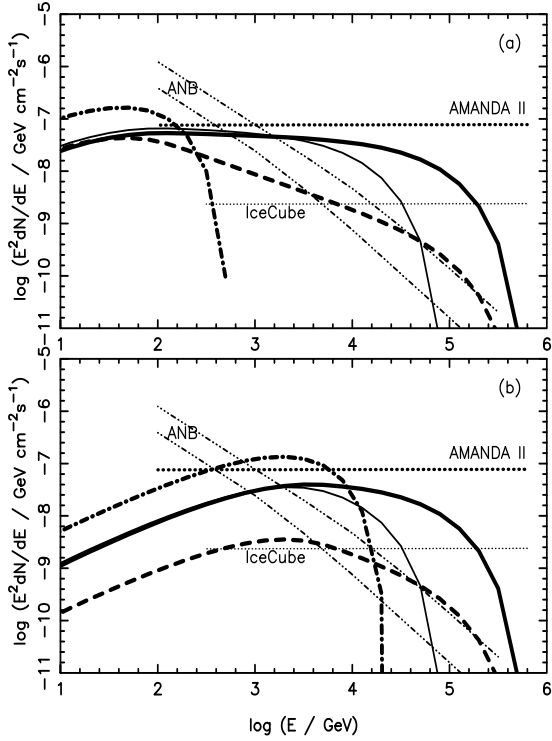

\vskip 7.truecm
\includegraphics{f2a.eps}
\includegraphics{f2b.eps}
\caption{\label{fig2} The differential energy spectra of muon neutrinos,
multiplied by $E^2$,  
produced in hadronic interactions of neutrons with the matter of the accretion disk
(a) and the massive Wolf Rayet star (b). The neutrino spectra are shown for the 
power law spectrum of 
nuclei, $\gamma_{\rm A}^{-\kappa}$, described by $\kappa = 2$ and the cut-off at 
$\gamma_{\rm A}^{\rm max}=10^7$ (thick full curves) and  at $10^6$ (thin full), and 
$\kappa = 2.5$ and  $\gamma_{\rm A}^{\rm max}=10^7$ (thick dashed), and  
for the mono-energetic nuclei with the Lorentz factor $2\times 10^5$ (accelerated 
in the region III) and $2\times 10^3$ (accelerated in the region II) (thick dot-dashed 
curves). The energy conversion efficiency from the jet to the relativistic nuclei is 
equal to $\chi = 10\%$. The dot-dot-dot-dashed curve indicate the 
atmospheric neutrino background within a $1^o$ of the source (Lipari 1993),
the thin dotted line shows the 3 yr sensitivity of the IceCube detector (Hill 2001),
and the thick dotted line shows the upper limit obtained by AMANDA II detector for 
the Cyg X-3 (Ackermann et al. 2004).}
\end{figure}


\begin{thebibliography}{}
\bibitem[Ackermann et al.~(2005)]{aketal05} Ackermann, M. et al. 2005, \prd, 
71, 077102
\bibitem[Aharonian \& Atoyan~(1991)]{aa91} Aharonian, F.A., Atoyan, A.M. 1991, \apj, 
381, 220
\bibitem[Anchordoqui et al.~(2003)]{anetal03} Anchordoqui, L.A., Torres, D.F., 
McCauley, T.P., Romero, F.A. 2003, \apj, 589, 481
\bibitem[Bartosik et al.~(2005)]{bbs05} Bartosik, M., Bednarek, W., Sierpowska, A. 
2003, in Proc. 28th ICRC (Tsukuba, Japan),  eds. Kajita et al. p. 2485  
\bibitem[Bednarek~(1993)]{bed93} Bednarek, W. 1993, A\&A, 278, 307
\bibitem[Bednarek~(2000)]{bed00} Bednarek, W. 2000, A\&A, 362, 646 
\bibitem[Berezinsly et al.~(1985)]{bcg85} Berezinsky, V.S., Castagnoli, C., 
Galeotti, P. 1985, Il Nuovo Cim., 8C, 185
\bibitem[Bishop~(1957)]{bi57} Bishop, G.R. 1957, Handbuch der Physik (Springer-Verlag), 
v. 42, p. 336
\bibitem[Bonnet-Bidaud \& Chardin~(1988)]{bc88} Bonnet-Bidaud, J.M., Chardin, G. 
1988, Phys.Rep., 170, 325
\bibitem[Carraminana~(1992)]{car92} Carraminana, A. 1992, A\&A, 264, 127
\bibitem[Cheng \& Ruderman~(1989)]{cr89} Cheng, K.S., Ruderman, M. 1989, \apj, 337, 
L77
\bibitem[Cheng \& Ruderman~(1991)]{cr91} Cheng, K.S., Ruderman, M. 1989, \apj, 373, 
187
\bibitem[Distefano et al.~(2002)]{dietal02} Distefano, C., Guetta, D., Waxman, E., 
Levinson, A. 2002, \apj, 575, 378
\bibitem[Falcke et al.~(2004)]{fal04} Falcke, H., K\"ording, E., Markoff, S. 2004,
A\&A, 414, 895 
\bibitem[Falcke et al.~(1995)]{fal95} Falcke, H., Malkan, M.A., Biermann, P.L. 1995, 
A\&A, 298, 375
\bibitem[Fender~(2003)]{fe03} Fender, R. 2003, in Compact stellar X-ray sources,
eds. W.H.G. Lewin \& M. van der Klis (Cambridge University Press) (astro-ph/0303339) 
\bibitem[Gaisser \& Grillo~(1987)]{gg87} Gaisser, T.K., Grillo, A.F. 1987, \prd, 36, 2752
\bibitem[Gaisser \& Stanev~(1985)]{gs85} Gaisser, T.K., Stanev, T., 1985, \prl, 
54, 2265 
\bibitem[Gandhi~(2000)]{ga00} Gandhi, R. 2000, Nucl.Phys.Suppl. 91, 453
\bibitem[Harding \& Gaisser (1990)]{hg90} Harding, A.K., Gaisser, T.K. 1990, \apj, 
358, 561
\bibitem[Heger \& Langer~(2000)]{hl00} Heger, A., Langer, N. 2000, \apj, 544, 1016 
\bibitem[Hill~(2001)]{hi01} Hill, G.C. 2001, in Proc. XXXVI Rencontres de Moriond,
(http://\-moriond.in2p3.fr/EW/2001/proceedings/) (astro-ph/0106064) 
\bibitem[Karaku\l a \& Tkaczyk~(1993)]{kt93} Karaku\l a, S., Tkaczyk, W. 1993, 
Astropart.Phys., 1, 229
\bibitem[Kolb et al.~(1985)]{ktw85} Kolb, E.W., Turner, M.S., Walker, T.P. 1985, 
\prd, 32, 1145
\bibitem[Levinson \& Waxman~(2001]{lw01} Levinson, A., Waxman, E. 2001, \prl, 87, 
171101
\bibitem[Lipari~(1993)]{lip93} Lipari, P. 1993, Astropart.Phys. 1, 195
\bibitem[Marti et al. (2001)]{mar01} Marti, J., Paredes, J.M., Peracaula, M. 2001, 
A\&A, 375, 476
\bibitem[Mioduszewski et al.~(2001)]{mio01} Mioduszewski, A.J. et al. 2001, \apj, 
553, 766
\bibitem[Mirabel \& Rodriquez~(1999)]{mr99} Mirabel, I.F., Rodriquez, L.F. 1999, 
ARA\&A, 37, 409
\bibitem[Nellen et al.~(1993)]{nmb93} Nellen, L., Mannheim, K., Biermann, P.L. 1993,
\prd, 47, 5270
\bibitem[Paredes~(2005)]{pa05} Paredes, J. M. 2005, in Int. Symp. on High Energy 
Gamma-ray Astronomy, eds. Aharonian et al., AIP, v. 745, p. 93 
%\bibitem[Romero \& Orllena~(2005)]{ro05} Romero, G.E., Orellana, M. 2005, A\&A, submitted
\bibitem[Romero et al.~(2003)]{roetal03} Romero, G.E., Torres, D.F., Kaufman Bernardo, 
M.M., Mirabel, I.F. 2003, A\&A, 410, L1
\bibitem[Sierpowska, A., Bednarek, W.~(2005)]{sb05} Sierpowska, A., Bednarek, W. 
2005, MNRAS, 356, 711 
\bibitem[Spencer et al.~(1986)]{sp86} Spencer, R.E., Swinney, R.W., Johnston, K.J., 
Hjellming, R.M. 1986, \apj, 309, 694
\bibitem[Van der Hucht~(2001)]{van01} Van der Hucht, K.A. 2001, New Astronomy, 45,
135 
\bibitem[Wdowczyk \& Wolfendale~(1987)]{ww87} Wdowczyk, J., Wolfendale, A.W. 1987, 
J.Phys.G., 13, 411
\bibitem[Yuan et al.~(2002)]{yuan02} Yuan, F., Markoff, S., Falcke, H.
2002, A\&A, 383, 854
\bibitem[Yuan et al.~(2002)]{yuanetal02} Yuan, F., Markoff, S., Falcke, H., 
Biermann, P.L. 2002, A\&A, 391, 139 

\end{thebibliography}
\end{document}